\documentclass[%
 reprint,
 amsmath,amssymb,
 aps,
]{revtex4-2}
\newcommand{\Int}{\int\limits}
\usepackage{graphicx}
\usepackage{dcolumn}
\usepackage{bm}
\usepackage{enumitem}
\usepackage{float}
\usepackage{mathtools}  
\usepackage{esvect} 
\usepackage{empheq}  
\usepackage{color}
\usepackage{caption}
\usepackage[most]{tcolorbox}  
\usepackage[makeroom]{cancel}  
\usepackage{nccmath}  
\usepackage{braket}
\usepackage{tikz} 
\usepackage{tabularx}				
\usepackage{array}					
\usepackage{hhline} 
\usepackage{physics}
\usepackage{siunitx}
\usepackage{hyperref}

\begin{document}

\preprint{APS/123-QED}

\title{Polarization entanglement with highly non-degenerate photon pairs\\enhanced by effective walk-off compensation method}

\author{Sungeun Oh\textsuperscript{a}}
\email{SungeunOh68@gmail.com}
\author{Thomas Jennewein\textsuperscript{a,b}}
\email{thomas.jennewein@uwaterloo.ca}
\affiliation{\textsuperscript{a}Institute for Quantum Computing, Department of Physics and Astronomy, University of Waterloo, Ontario, Canada N2L 3G1 \\ 
\textsuperscript{b}Department of Physics, Simon Fraser University, Burnaby, British Columbia, Canada V5A1S6}

\date{\today}

\begin{abstract}
We present an entangled photon source (EPS) that produces polarization entanglement in highly non-degenerate photon pairs, generated through Type-0 spontaneous parametric down conversion (SPDC) using bulk periodically poled Lithium Niobate (PPLN). This source is specifically designed for  entanglement-based quantum key distribution (EBQKD) between ground and satellite, as part of the Quantum Encryption and Science Satellite (QEYSSat) mission funded by the Canadian Space Agency (CSA). Through the utilization of birefringent beam displacers in a Sagnac interferometer scheme, we ensure high polarization contrast and stable interference of highly non-degenerate photon pairs. However, this approach causes substantial spatial and temporal walk-offs of the photon paths, posing a formidable challenge. We introduce an effective compensation method using birefringent crystal wedges to eliminate spatial and temporal walk-offs simultaneously. With only $\SI{1.0}{\milli\watt}$ of pump power, we observed a coincidence rate of $(33.33\pm0.05)$kHz, a significant improvement compared to the rate of $(4.20\pm0.01)$kHz without spatial compensation. Additionally, we observed an estimated pair generation rate of $(2.92\pm0.12)$MHz and an entanglement visibility of $(97.1\pm0.3)\%$, making this a promising source for long-distance quantum communication in ground-to-satellite and fiber optic links.
\end{abstract}

\maketitle

\section{Introduction}

While numerous source for polarization entanglement with degenerate or near-degenerate photon pairs have been demonstrated \cite{tang2015photon,PhysRevA.73.012316,PhysRevA.77.032338,PhysRevApplied.18.044027}, polarization entanglement with highly non-degenerate photon pairs has received less attention. However, sources that produce highly non-degenerate photon pairs are relevant\cite{PhysRevApplied.18.044075,Sidhu_2021,Ursin_2007,Hentschel_2009} in the context of quantum communication scenarios that require distinct wavelengths. Of our particular interest is the establishment of ground-to-space entanglement-based quantum key distribution (EBQKD)\cite{Li_2023,PhysRevA.76.012307}, which necessitates a high pair production rate of around 100~MHz\cite{Bourgoin_2013}, as well as the utilization of connectivity to both the satellite free-space link and the ground-based optical fiber link. The photon transmitted through the optical fiber should preferably have a wavelength around $\SI{1550}{\nano\meter}$ to ensure compatibility with existing optical fiber infrastructure. The ground-to-space link will operate with a wavelength in the visible or near-infrared range which is optimal for free-space communication \cite{Bourgoin_2013}. However, deploying SNSPDs on satellites is challenging due their cryogenic cooling requirements \cite{10.1063/1.5000001}. The Silicon-based photon detectors excel in performance and have shown reliable operation in space environments \cite{Yang:24,10.1029/2005GL024028,DSouza_2021,PhysRevApplied.20.044052}. Among these, Silicon-based single-photon avalanche diodes (Si-SPADs) are chosen for quantum receivers in the QEYSsat payload \cite{Bourgoin_2013}. 

Under these circumstances, the photon pairs used for the ground-to-space QKD are highly degenerate in wavelength. We have outlined the challenges arising from this high degeneracy and present the design and techniques employed to overcome these difficulties. We present an entangled photon source (EPS) specifically designed for implementing EBQKD between ground-based and satellite-based systems, as part of the Quantum Encryption and Science Satellite (QEYSSat) mission funded by the Canadian Space Agency (CSA)\cite{QEYSSatproposal,Pugh_2017,QEYSSatMission}.
\section{Scheme to Generate Polarization Entangled non-degenerate Photon Pairs}

Our EPS generates photon pairs through spontaneous parametric down conversion (SPDC) \cite{Steinlechner_2012} in optically nonlinear crystals. A continuous-wave (CW) laser (Frankfurt laser, Germany) provides a polarized pump beam at a wavelength of $\lambda_p=\SI{523.6}{\nano\meter}$, with a bandwidth of $\Delta\lambda_p=\SI{0.1}{\nano\meter}$. A $\SI{10}{\milli\meter}$ long Type-0 PPLN crystal (Covesion, UK) with a poling period of $\Lambda=\SI{7.1}{\micro\meter}$ (at room temperature) converts pump(p) photons into signal(s)-idler(i) photon pairs with the same polarization under quasi-phase matching condition.
\begin{equation}
\setlength{\jot}{10pt}  
	\begin{aligned}\label{QPM}
	\frac{2\pi}{\Lambda(T)}=k_p-k_s-k_i
	\end{aligned}
\end{equation}
$k=2\pi n(\lambda,T)/\lambda$ represent the wave vectors, where $n(\lambda,T)$ denote the wavelength- and temperature-dependent refractive indices of the crystals for the corresponding fields. We chose this crystal for its higher photon pair generation rate compared to other nonlinear crystals \cite{2008ApPhB}. With the presented values, we have signal photons at a wavelength of $\lambda_s=790.8nm$ suitable for free-space transmission, and idler photons at a wavelength of $\lambda_i=1550nm$ suitable for ground transmission through optical fiber. The associated spectral bandwidths $\Delta\omega$ (at 1/e) are estimated using the first-order Taylor expansion of equation \ref{QPM}.
\begin{equation}
\setlength{\jot}{10pt}  
	\begin{aligned}\label{Bandwidth}
	\Delta k\approx\left.\dfrac{\partial k_{p}}{\partial \omega}\right|_{\omega_{p}}\Delta\omega_{p}-\left.\dfrac{\partial k_{s}}{\partial \omega}\right|_{\omega_{s}}\Delta\omega_{s}-\left.\dfrac{\partial k_{i}}{\partial \omega}\right|_{\omega_{i}}\Delta\omega_{s}=\dfrac{2\pi}{L}
	\end{aligned}
\end{equation}
Using equation \ref{Bandwidth}, the temporal widths (at 1/e) are approximately $\tau_s=\SI{2.68}{\pico\second}$ for the signal photon, and $\tau_i=\SI{2.61}{\pico\second}$ for the idler photon. We consider the entanglement of signal and idler photons with horizontal ($H$) and vertical ($V$) polarizations using an optical interferometer. The resulting Bell state can be written as
\begin{equation}
\setlength{\jot}{10pt}  	\begin{aligned}\label{BellState}	\ket{\text{Bell}}=\dfrac{1}{\sqrt{2}}\left(\ket{V_sV_i}+e^{i\left(\phi_r+\Delta\phi_r\right)}\ket{H_sH_i}\right)
	\end{aligned}
\end{equation}
The overall relative phase between the $\ket{VV}$ and the $\ket{HH}$ states is the result of a combination of the relative phase due to the path difference (denoted as $\phi_r$) and the relative phase variation induced by temperature fluctuations in optical components (denoted as $\Delta\phi_r$). One interferometer design widely used for generating entangled photon pairs is the Sagnac interferometer, which incorporates self-compensation for relative phase stabilization \cite{Hentschel_2009}. However, a Sagnac interferometer with a polarization beam splitter, which is commonly used, has technical difficulties in achieving entanglement in non-degenerate photon pairs with high-efficiency transmission for both signal and idler photons, while simultaneously accommodating their two orthogonal polarizations. The reason lies in the wavelength dependency of the transmittivity and reflectivity ratio of the polarization beam splitter \cite{makarov2022theory}. 

To overcome this challenge, two birefringent beam displacers capable of splitting and recombining two orthogonally polarized beams are used to build an optical interferometer for the EPS. A half waveplate (HWP) rotated by $45^o$ is inserted between the two beam displacers, such that the polarizations of the ordinary ray (non-deflected, denoted with subscript '$o$') and the extraordinary ray (deflected ray, denoted with subscript '$e$') at the first beam displacer switch roles by the time they reach the second beam displacer. This ensures equal path lengths \cite{Shi_2020,lee2021sagnac,szlachetka2022ultrabrightsagnactypesourcenondegenerate,Lohrmann_2020}. 
\begin{figure}[htbp!]
\centering
\includegraphics[width=\columnwidth]{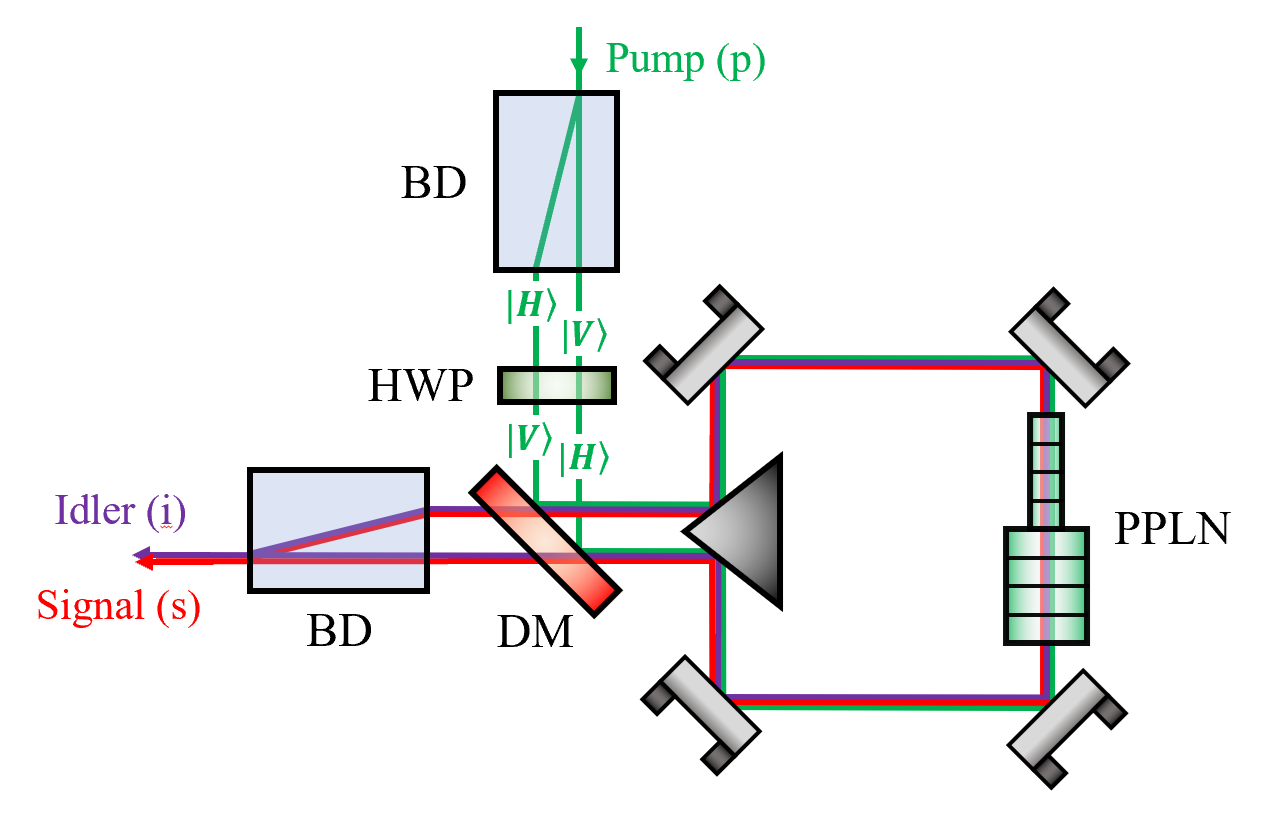}
\footnotesize
\caption{Schematic diagram of the interferometer implementing two beam displacers and a half waveplate and including Sagnac loop. Two PPLN crystals are placed in the loop, each responsible for the SPDC of one of the polarized pump beams.}\label{Interferometer}
\end{figure}
This scheme avoids the problem associated with the single polarization beam splitter scheme and allows for the effective transmission of all three waves through the interferometer. 

The perfect alignment of the two beam displacers can be seen as an equivalent configuration to the Sagnac interferometer, but this is always an idealized case. We construct an actual Sagnac loop into the interferometer by inserting a dichroic mirror along the beam path between the half waveplate and the second beam displacer. The dichroic mirror reflects the two orthogonally polarized pump beams, allowing them to enter the Sagnac loop, and transmits the signal and idler beams with both polarizations effectively. This Sagnac loop also prevents back reflection from the interferometer to the pump laser. The returning pump beam which has not undergone down-conversion could otherwise be recombined at the first beam displacer along the same pump beam path, potentially causing laser instability or even damage. The phase variations of each of the separated beam paths along the Sagnac loop (denoted with subscripts, CW for clockwise and CCW for counter clockwise) are
\begin{equation}
\setlength{\jot}{10pt}  
	\begin{aligned}\label{PathPhase1}
	\Delta\phi_{CW}=\delta\phi_{e}(\lambda_p,T)+\delta\phi_o(\lambda_s,T)+\delta\phi_o(\lambda_i,T)\\
	\Delta\phi_{CCW}=\delta\phi_o(\lambda_p,T)+\delta\phi_{e}(\lambda_s,T)+\delta\phi_{e}(\lambda_i,T)\\
	\end{aligned}
\end{equation}
where we have
\begin{equation}
\setlength{\jot}{10pt}  
\begin{aligned}\label{PhaseVariation}	\delta\phi=\sum_j\dfrac{2\pi L_j}{\lambda}\left(\dfrac{\partial n_{j}}{\partial T}+n_{j}\alpha_{j}\right)\Delta T,
	\end{aligned}
\end{equation}
representing the phase shift introduced by all transmissive optical components, each indexed with $j$, within the interferometer, each with a length of $L_j$. The thermo-optic coefficients $\partial n_{j}/\partial T$ and the thermal expansion coefficient $\alpha_{j}$ in equation \ref{PhaseVariation} correspond to changes in the refractive index and the material length, respectively. 
The relative phase variation is then
\begin{equation}
\setlength{\jot}{10pt}  
	\begin{aligned}\label{PathPhase3}
	\Delta\phi_r&=\Delta\phi_{CW}-\Delta\phi_{CCW}\\
	&=\delta\phi_r(\lambda_p,T)-\delta\phi_r(\lambda_s,T)-\delta\phi_r(\lambda_i,T)
	\end{aligned}
\end{equation}
where $\delta\phi_r=\delta\phi_e-\delta\phi_o$. We assume that the phases introduced by the half waveplate and the dichroic mirror are negligible compared to those induced by the beam displacers, due to their much smaller thicknesses. Phase variations are then calculated to be $\delta\phi_r(\lambda_p,T)=6.22\pi$, $\delta\phi_r(\lambda_s,T)=4.09\pi$ and $\delta\phi_r(\lambda_i,T)=2.08\pi$. Hence, the relative phase is $\Delta\phi_r=0.052\pi$, ensuring sufficient phase stability. No self-compensation effect would introduce a phase shift of $\Delta\phi_r=12.39\pi$. As a result, while we don't achieve a perfect self-compensation, it certainly contributes to maintaining the stability of quantum states over a longer period of time. 

\section{Spatiotemporal Walk-off Compensation}

Even if the relative phase ($\phi_r$ in equation \ref{BellState}) is minimized to some extent by the Sagnac loop of our source, there is still phase introduced by the two beam displacers. 
The amount of displacement of the extraordinary ray in the beam displacers can be determined from its angle of refraction, denoted as $\theta_e$. This angle depends on the refractive indices of the ordinary and extraordinary rays (denoted as $n_o$ and $n_e$, respectively).
\begin{equation}
\setlength{\jot}{10pt}  
	\begin{aligned}\label{AOR}
	\tan{\theta_e}=\left(1-\dfrac{n_o^2}{n_e^2}\right)\dfrac{\tan{\theta}}{1+\dfrac{n_o^2}{n_e^2}\tan^2{\theta}}
	\end{aligned}
\end{equation}
The optic angle (denoted as $\theta$) is the angle between the optic axis and the axis normal to the incidence surface of the medium.
It is worth noting that equation \ref{AOR} assumes the incident beam entering the medium perpendicular to the surface of the medium. The separation between the two rays created by the beam displacer with a length of $L$ is then $D=L\tan{\theta_e}$. 
\begin{figure}[htbp!]
\centering
\includegraphics[width=\columnwidth]{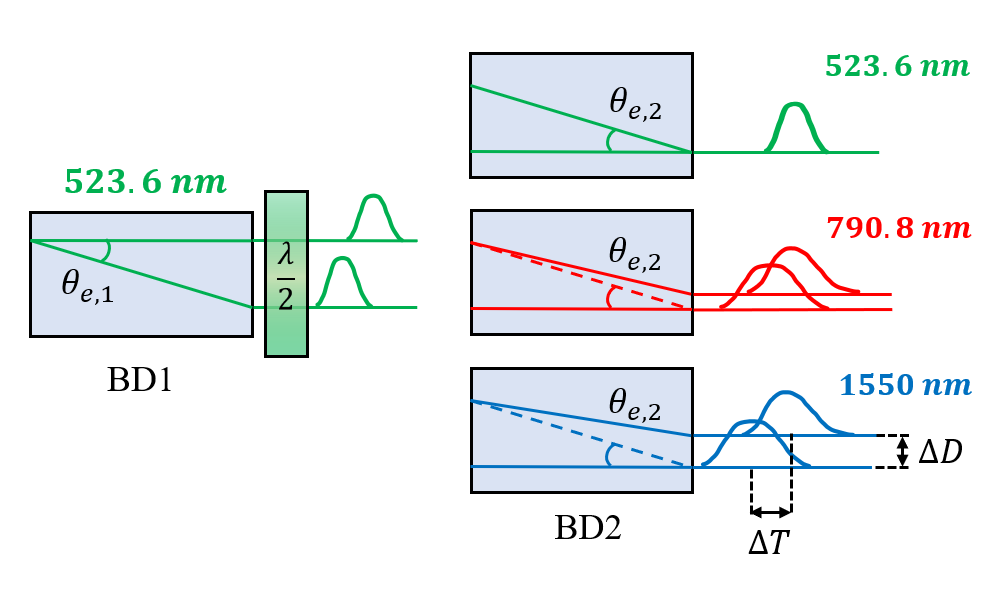}
\footnotesize
\caption{Diagrams illustrating how the signal and idler beams undergo spatial $\Delta D$ and temporal $\Delta T$ walk-offs in the second beam displacer. If the beams had maintained the pump wavelength, then no walk-off would have occurred at the second beam displacer. The degree of displacement decreases as the wavelength increases.}\label{Walkoff}
\end{figure}
When the down-converted signal and idler photons pass through the second beam displacer, they do not return to their original position, as their wavelengths are no longer the same as the pump wavelength (see Figure \ref{Walkoff}). Spatial walk-off reduces the entangled photon pair rate due to the misalignment in fiber-coupling, while temporal walk-off degrades the entanglement visibility by reducing the temporal overlap of two wavepackets. Both effects worsen with more extreme degenerate photon pairs. The spatial walk-off, denoted as $\Delta D$, is the deviation of the e-ray from its original path.
\begin{equation}
\setlength{\jot}{10pt}  
	\begin{aligned}\label{SWalkoff}
	\Delta D_{s,i}=L\tan{\theta_{e,2}^p}-L\tan{\theta_{e,1}^{s,i}}
	\end{aligned}
\end{equation}
$\theta_{e,1}$ and $\theta_{e,2}$ are the refraction angles of the first and the second beam displacers, respectively, as shown in Figure \ref{Walkoff}. We use two identical $\alpha$BBO beam displacers (Newlight Photonics inc, Canada) with lengths of $L=\SI{39.4}{\milli\meter}$ for the interferometer. According to equation \ref{SWalkoff}, the spatial walk-offs of the signal and the idler beams are $\SI{0.10}{\milli\meter}$ and $\SI{0.17}{\milli\meter}$, respectively. The time taken for the beam to pass through the beam displacer is described using a geometric approach \cite{lee2021sagnac}. 
\begin{equation}
\setlength{\jot}{10pt}  
	\begin{aligned}\label{TimeTraveled}
	T_{o\rightarrow e}^{s,i}=&\dfrac{L}{v_{g,o}^p}+\dfrac{L}{v_{g,e}^{s,i}\cos{\theta_{e,2}^{s,i}}}\\
	T_{e\rightarrow o}^{s,i}=&\dfrac{L}{v_{g,e}^p\cos{\theta_{e,1}^p}}+\dfrac{L}{v_{g,o}^{s,i}}
	\end{aligned}
\end{equation}
$T_{o\rightarrow e}$ is the time taken by the ray that initially travels straight and then becomes displaced. Similarly, $T_{e\rightarrow o}$ is the time taken by the ray that initially becomes displaced and then travels straight. $v_{g,o}$ and $v_{g,e}$ are the group velocities of the o-ray and e-ray, respectively. The temporal walk-off is then the time difference between these two travel times.
\begin{equation}
\setlength{\jot}{10pt}  
	\begin{aligned}\label{TWalkoff}
	\Delta T_{s,i}=T_{o\rightarrow e}^{s,i}-T_{e\rightarrow o}^{s,i}
	\end{aligned}
\end{equation}
Using equation \ref{TWalkoff}, we obtain temporal walk-offs of $\SI{0.65}{\pico\second}$ for the signal beam and $\SI{1.06}{\pico\second}$ for the idler beam. 
Employing birefringent materials to compensate for only temporal walk-offs is a commonly used method. Effectively compensating for both spatial and temporal walk-off simultaneously has been a challenging problem. One approach could be using two separate beam displacers for the signal and idler beams to eliminate spatial walk-off \cite{PhysRevApplied.18.044075}. However, this would require an additional half waveplate to fine-tune the temporal overlap, thereby increasing the size and complexity of the setup. 

Here, we employ pairs of birefringent crystal wedges with adjustable separation and thickness to achieve simultaneous compensation for both spatial and temporal walk-offs. 
\begin{figure}[htbp!]
\centering
\includegraphics[width=\columnwidth]{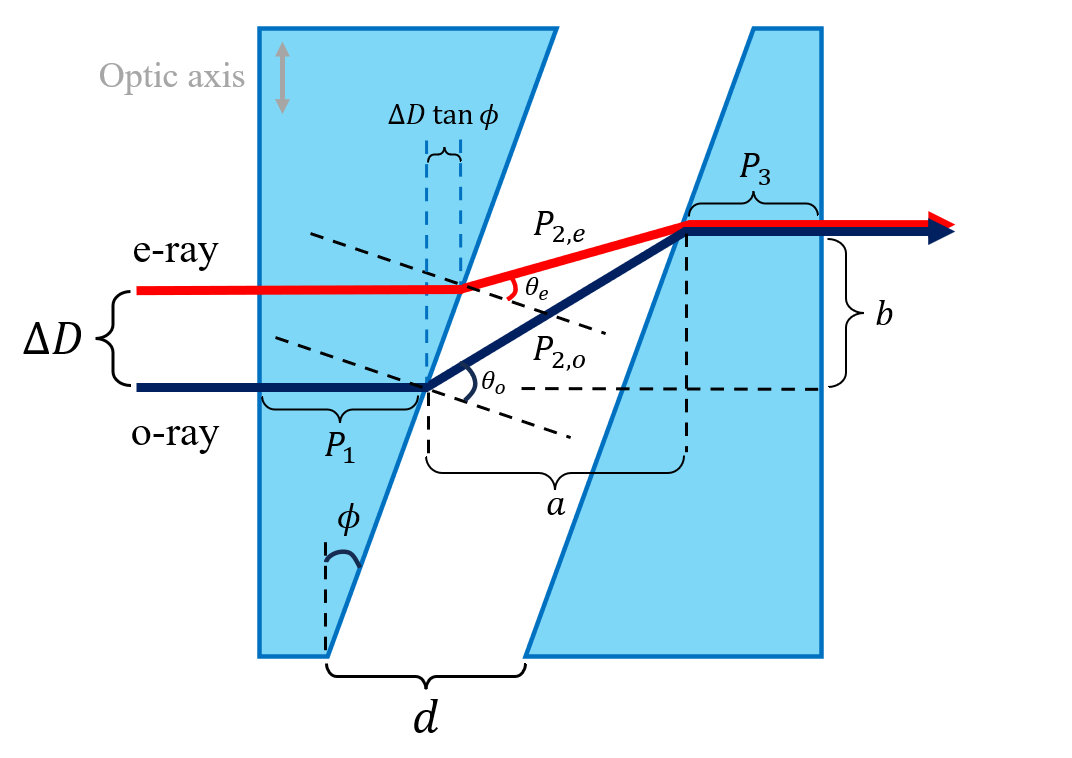}
\footnotesize
\caption{Schematic diagram of two Calcite wedges used to compensate for both spatial and temporal walk-offs. The optic axis of our Calcite wedges is aligned perpendicular to the direction of ray propagation such that both the ordinary (purple) and extraordinary (red) rays undergo no beam displacement when entering the Calcite perpendicularly. However, upon exiting the first wedge, the e-ray experiences more refraction than the o-ray due to negative birefringence ($n_e<n_o$). The path lengths of the e-ray and the o-ray in between the two wedges are shown as $P_{2,e}$ and $P_{2,o}$, respectively.}\label{Compensation}
\end{figure}
The choice of a wedge shape offers several advantages. As illustrated in Figure \ref{Compensation}, the incidence plane is perpendicular to the incidence rays, simplifying the calculations for group velocities appearing in equation \ref{TimeTraveled}. Moreover, even if the ray exits the first wedge at a non-perpendicular angle, it regains its original propagation direction when it enters the second wedge that is identical in shape. By identifying the correct lateral separation (shown as $d$ in Figure \ref{Compensation}) between the two wedges, we effectively merge the two rays at the second wedge, compensating for the spatial walk-offs. At the same time, temporal walk-offs can easily be compensated for by adjusting the thickness of the wedges. We use Calcite as birefringent wedges with a wedge angle of $\phi=\SI{15}{\degree}$. Since Calcite is a negative uniaxial material, o-rays experience more refraction than the e-rays in both the signal and the idler path. It is worth noting that in the first Calcite wedge (on the left in Figure\ref{Compensation}), the e-ray travels a slightly longer distance than the o-ray by $\Delta D\tan{\phi}$. Additionally, when a gap between the two crystal wedges is considered, the distance the two beams travel within the gap differs slightly due to the different bending of their paths. Consequently, the o-ray travels a longer distance than the e-ray, resulting in a reduction of the temporal walk-off. This leads to either an increase or a decrease in the initial temporal walk-off, and all of these should be taken into account when determining the thickness of the wedges. 
The vertical displacement of the o-ray is $b=a\tan{\left(\theta_o-\phi\right)}$ such that the vertical shift of the e-ray is $b-\Delta D$. The relationship between the two vertical displacements is as follows.
\begin{equation}
\setlength{\jot}{10pt}  
	\begin{aligned}\label{Compensation1}
	a\tan{\left(\theta_o-\phi\right)}-\Delta D=\left(a-\Delta D\tan{\phi}\right)\tan{\left(\theta_e-\phi\right)}
	\end{aligned}
\end{equation}
$\theta_{o,e}$, $\phi$ and $\Delta D$ are known quantities, so we can rearrange equation \ref{Compensation1} to determine $a$.
\begin{equation}
\setlength{\jot}{10pt}  
	\begin{aligned}\label{Compensation2}
	a=\dfrac{1-\tan{\phi}\tan{\left(\theta_e-\phi\right)}}{\tan{\left(\theta_o-\phi\right)}-\tan{\left(\theta_e-\phi\right)}}\Delta D
	\end{aligned}
\end{equation}
Finally, we calculate the lateral separation $d$ as follows.
\begin{equation}
\setlength{\jot}{10pt}  
	\begin{aligned}\label{Compensation3}
	d=a-b\tan{\phi}
	\end{aligned}
\end{equation}
The total path lengths for each ray are expressed with reference to Figure \ref{Compensation}.
\begin{equation}
\setlength{\jot}{10pt}  
	\begin{aligned}\label{Compensation4}
	P(\text{o-ray})&=P_1+P_{2,o}+P_3\\
	P(\text{e-ray})&=P_1+\Delta D\tan{\phi}+P_{2,e}+P_3
	\end{aligned}
\end{equation}
The difference in travel times between the two paths is then
\begin{equation}
\setlength{\jot}{10pt}  
	\begin{aligned}\label{Compensation5}
	\Delta T&=\left[\dfrac{1}{v_{g,o}}\left(P_1+P_3\right)+\dfrac{1}{c}P_{2,o}\right]\\&-\left[\dfrac{1}{v_{g,e}}\left(P_1+\Delta D\tan{\phi}+P_3\right)+\dfrac{1}{c}P_{2,e}\right]
	\end{aligned}
\end{equation}
By setting the overall thickness of the wedges to $L=\left|v_{g,e}-v_{g,o}\right|\Delta T$, we effectively cancel out the temporal walk-off. Gaussian fields are considered for characterizing the spatial modes of the ordinary and extraordinary rays with their beam widths, $\sigma$.
\begin{equation}
\setlength{\jot}{10pt}  
	\begin{aligned}\label{Gaussian}
	g(x,y)=\dfrac{1}{2\pi\sigma^2}e^{-(x^2+y^2)/2\sigma^2}
	\end{aligned}
\end{equation}\\
The sptial overlap of the two beam separated by $\Delta D$ is then the product of the two Gaussians integrated over the transverse plane (x,y).
\begin{equation}
\setlength{\jot}{10pt}  
	\begin{aligned}\label{BeamOverlap}
	\Int_{-\infty}^{\infty}g_o(x,y)g_e(x-\Delta D_x,y-\Delta D_y)dxdy
	\end{aligned}
\end{equation}
The impact of spatial walk-offs is examined by multiplying the ideal photon rate with the spatial overlap. The theoretical curve of the spatial overlap is shown on Figure \ref{CompensationResults}b. The temporal overlap can also be examined using equation \ref{Gaussian}, considering temporal modes of the ordinary and extraordinary rays instead of the spatial modes. However, it is worth noting that SPDC occurs randomly in time when a CW laser produces the pump field. Therefore, in this experiment, we focus on minimizing the relative temporal walk-off $\left|\Delta T_s-\Delta T_i\right|$ between the signal and the idler beams, rather than minimizing them individually. 
\section{Experiments and Results}

\begin{figure}[htbp!]
\centering
\includegraphics[width=\columnwidth]{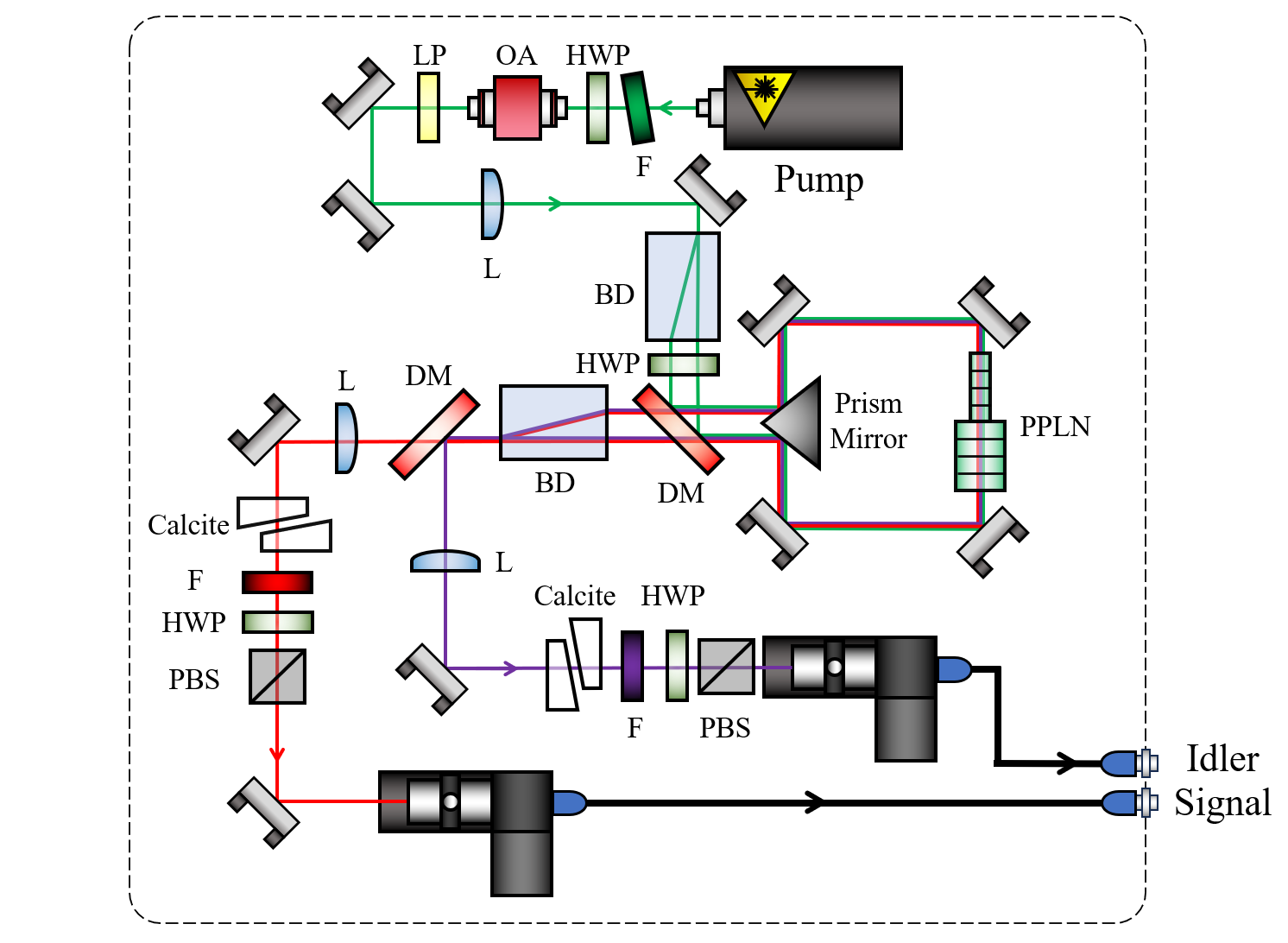}
\footnotesize
\caption{Design of the EPS. The optical components are : Filter (F), Optical Isolator (OA), Linear Polarizer (LP), Lens (L), Beam Displacer (BD), Half waveplate (HWP), Dichroic Mirror (DM), Polarizing Beam splitter (PBS)}\label{EPS}
\end{figure}

We demonstrate polarization-entanglement of photons with our EPS which implements the hybrid interferometer scheme (as shown in Figure \ref{EPS}). 
The pump beam collimated by an aspheric lens ($f_p=\SI{8}{\milli\meter}$) passes through several optical components to be diagonally polarized. It is then focused by a magnifying lens ($f_p=\SI{40}{\centi\meter}$) onto the PPLN crystals in the Sagnac loop. The first beam displacer splits the diagonally polarized pump beam into two beams with horizontal and vertical polarization by displacing one of them, which is extraordinary with respect to the displacer's optic axis. The polarization of the two beams is switched by passing through a half waveplate before they enter the Sagnac loop. The first dichroic mirror reflects the pump beam into the Sagnac loop, while allowing the signal and idler beams produced from down-conversion by the PPLNs to pass through and reach the second beam displacer. We place the two identical PPLNs perpendicular to each other, with one responsible for $\ket{o}_p\rightarrow\ket{o}_s\ket{o}_i$ and the other responsible for $\ket{e}_p\rightarrow\ket{e}_s\ket{e}_i$. The two signal and idler beams combined by the second beam displacer are separated by the second dichroic mirror (the signal beam is transmitted and the idler beam is reflected). Along each beam path, another lens ($f_s=\SI{60}{\centi\meter}$ and $f_i=\SI{50}{\centi\meter}$) is placed to re-collimate the beam before entering the Calcite wedges for walk-off compensation. 
\begin{figure}[htbp!]
\centering
\includegraphics[width=\columnwidth]{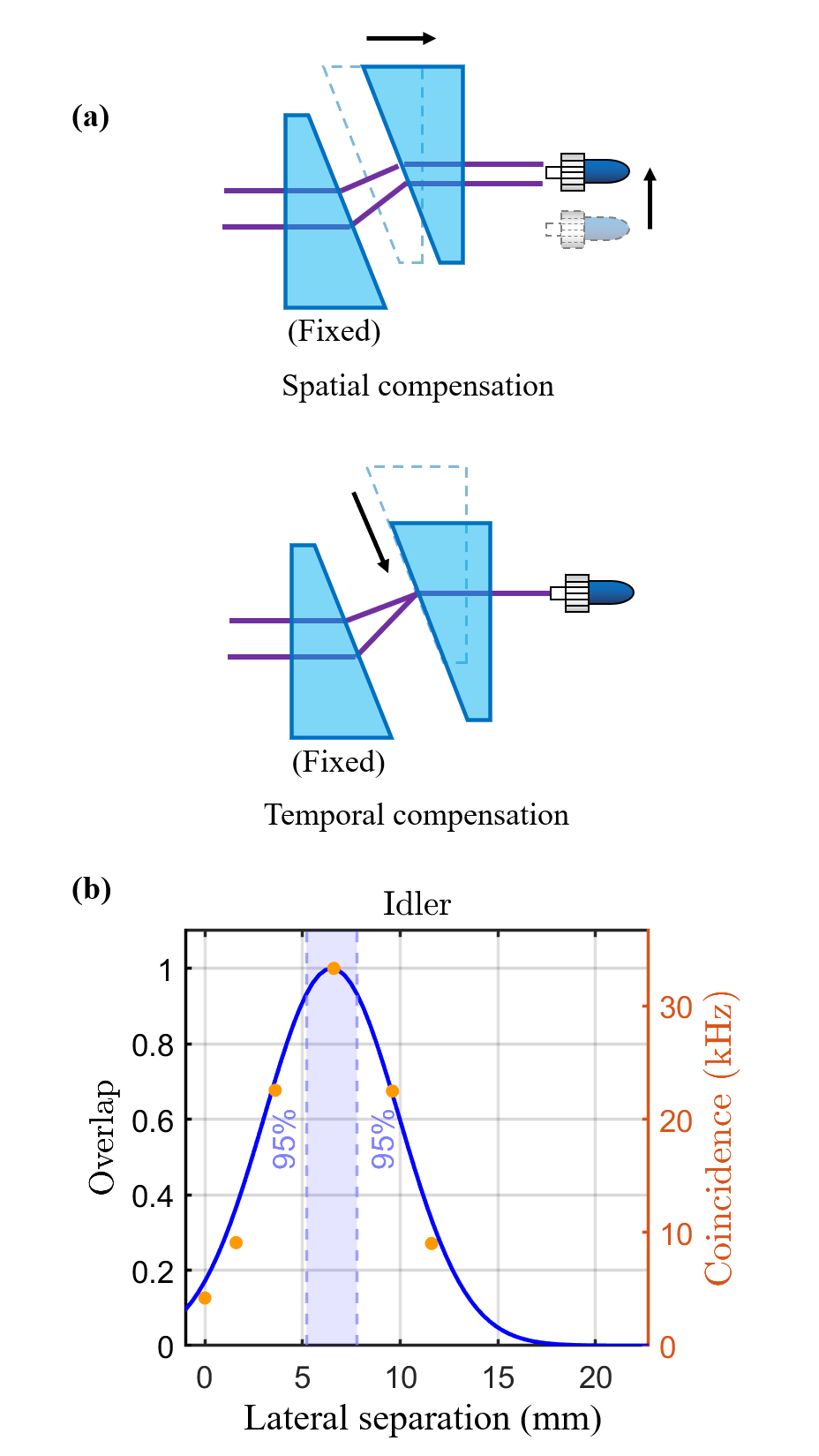}
\footnotesize
\caption{(a) Simultaneous walk-off compensation. Translation of one of the wedges along the beam propagation direction allows compensation for the spatial walk-off. The optical fiber collecting idler photons is realigned in order to maximize the coupling efficiency. The thickness is varied by the diagonal translation of the same wedge, allowing compensation for the temporal walk-off. (b) Theoretical curve of the spatial overlap for idler photons (solid line) as a function of lateral separation, along with the calculated coincidences (dot) for each lateral separation. Note that both the vertical and horizontal error bars are too short to be visible.
}\label{CompensationResults}
\end{figure}
This prevents the beam from entering non-perpendicular to the wedges' surfaces. 

In actual experiments, higher values of $\Delta D_s=\SI{0.145}{\milli\meter}$ and $\Delta D_i=\SI{0.325}{\milli\meter}$ are observed due to additional divergence at the two dichroic mirrors. Nevertheless, compensating for the spatial walk-off should pose no issues with the additional divergence and should simply be taken into account. The FWHM of the signal and idler beams at the receiving fibers are $\SI{0.6}{\milli\meter}$ and $\SI{0.8}{\milli\meter}$, respectively, as determined through the analysis of their intensity profiles. With the actual walk-offs obtained from the experiment, the spatial overlap factors calculated from equation \ref{BeamOverlap} are 52.4\% for the signal photon, and 17.1\% for the idler photon. Using equation \ref{Compensation3}, the lateral separation needed for the signal and idler paths are calculated to be $d_s=\SI{2.75}{\milli\meter}$ and $d_i=\SI{6.6}{\milli\meter}$, respectively. Since Calcites are placed in both beam paths for spatial compensation, we fix the thickness of the Calcite wedges in the signal path and adjust the temporal walk-off of the idler beam accordingly. The adjustment can be done simply by allowing diagonal translation of the idler Calcite wedge as shown in Figure \ref{CompensationResults}a. One would expect the signal and idler photons to be completely entangled after the wedges. A long-pass filter and a bandpass filter are placed on each beam path after the wedges to reduce any spectral noise. The polarization-entanglement test is proceeded by measuring the polarization correlation of the Bell states. A half-wave plate and a polarization beam splitter are placed at the end of each beam path (signal and idler). The arrived signal and idler photons are focused by the aspheric lenses ($f_s=\SI{11}{\milli\meter}$ and $f_i=\SI{15}{\milli\meter}$) and collected through the single-mode fibers. Si-SPAD (Excelitas, Canada) and SNSPD (Quantum Opus, USA) are used to collect signal and idler photons, respectively. The detection efficiencies for the signal and the idler beam paths are estimated to be $\eta_{s}=0.38$ and $\eta_{i}=0.217$ in total, considering all the optical components along the ways including the detectors, assuming efficiencies based on the manufacturers' specifications. The coincidence counts within a $\SI{1.5}{\nano\second}$ detection window are measured using a time-tagging unit (UQDevices, CA). 

To validate our compensation for the spatial walk-off, the lateral separation between the wedges on the idler side is varied to $\SI{0}{\milli\meter}$, $\SI{1.6}{\milli\meter}$, $\SI{3.6}{\milli\meter}$, $\SI{6.6}{\milli\meter}$, $\SI{9.6}{\milli\meter}$ and $\SI{11.6}{\milli\meter}$, while the spatial walk-off on the signal side is fully compensated (see Figure \ref{CompensationResults}a). With the pump power of $\SI{1.0}{\milli\watt}$, background-subtracted coincidence counts are measured for each variation and plotted on Figure \ref{CompensationResults}b. The peak experimental value on the idler graph corresponds to when both spatial and temporal walk-offs are fully compensated for both the signal and idler beams, with the coincidence counts measuring $N=(33.33\pm0.05)$kHz. This marks almost eight times enhancement in the coincidence counts, as without compensation on the idler side (while the signal side remained fully compensated), the coincidence count was $N=(4.20\pm0.01)$kHz.
The correlations are measured by rotating the half waveplate on the signal arm while the half waveplate on the idler arm is fixed to $0^o$ (V basis), $45^o$ (A basis), $90^o$ (H basis) and $135^o$ (D basis). Then, the visibility $V=(N_{c,max}-N_{c,min})/(N_{c,max}+N_{c,min})$ of each of the four bases is obtained from the minimum and maximum coincidence counts of the correlation measurements. A peak average value of $V=(97.1\pm0.3)\%$ was obtained by tuning the thickness of the Calcite wedges. Furthermore, based on the compensation results, the additional beam divergence from the dichroic mirrors appears to have had minimal, if any, impact on the temporal walk-off compensation, as expected, since we only had to minimize the relative temporal walk-offs, as mentioned earlier.
\begin{figure}[htbp!]
\centering
\includegraphics[width=\columnwidth]{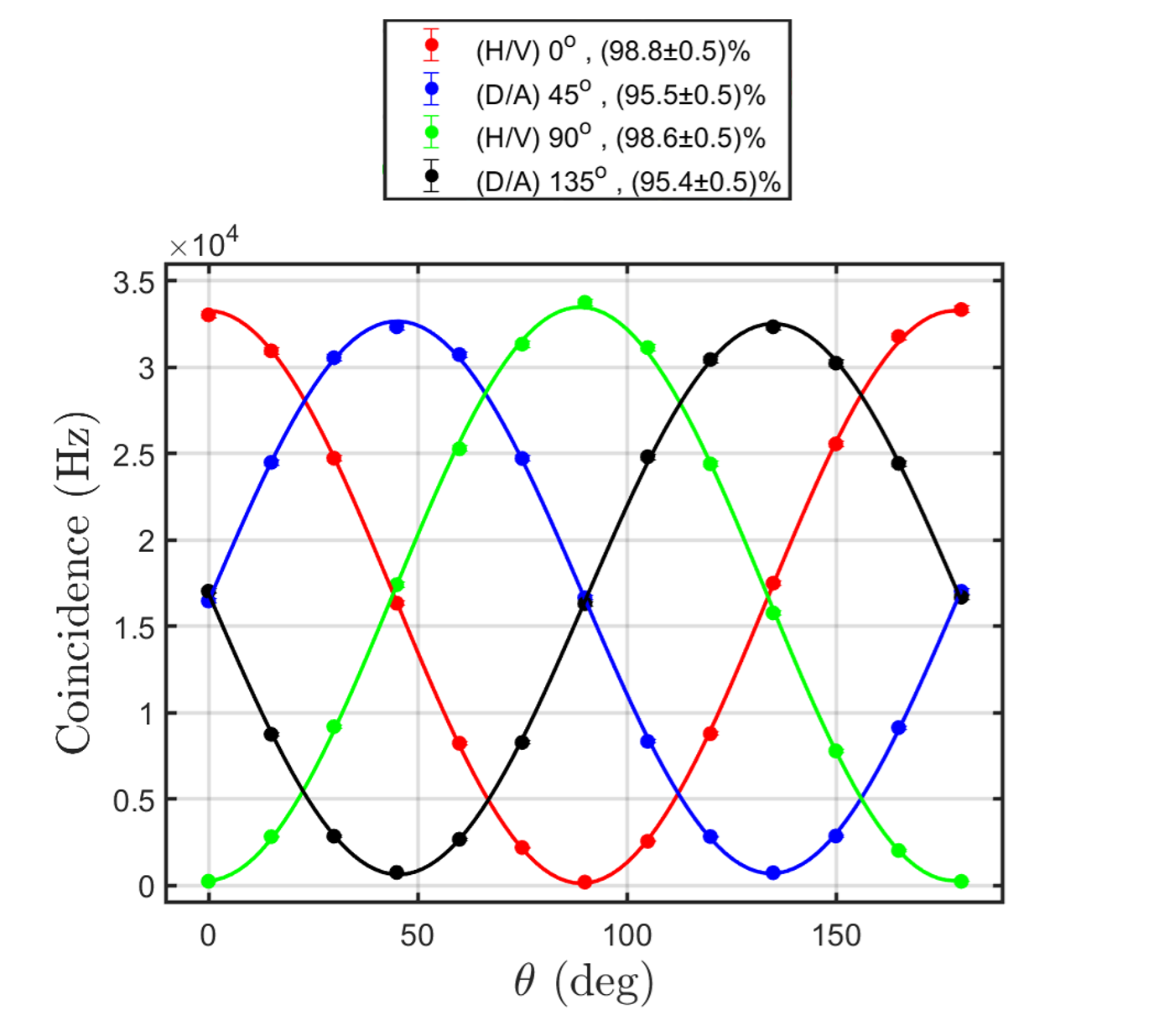}
\footnotesize\caption{Polarization entanglement measurement results. The correlation measurement was conducted using four different polarization bases. The coincidence counts, measured at $\SI{1.0}{\milli\watt}$ of pump power while changing the polarization angle from $0^o$ to $180^o$, are plotted on the graph.}\label{UltimateResult}
\end{figure}
Figure \ref{UltimateResult} shows the optimal entanglement test results. The measured visibilities with full walk-off compensation are $V=(98.8\pm0.5)\%$, $V=(95.5\pm0.5)\%$, $V=(98.6\pm0.5)\%$ and $V=(95.4\pm0.5)\%$. In order to confirm the entanglement, we perform the CHSH test \cite{PhysRevLett.23.880}. The CHSH inequality is expressed as 
\begin{equation}
\setlength{\jot}{10pt}  
	\begin{aligned}\label{CHSH}
	S=\left|E_{\alpha,\beta}-E_{\alpha,\delta}+E_{\gamma,\delta}+E_{\gamma,\beta}\right|\le2
	\end{aligned}
\end{equation}
where we have the correlation estimates expressed as
\begin{equation}
\setlength{\jot}{10pt}  
	\begin{aligned}\label{CHSHest}	E_{\alpha,\beta}=\dfrac{N\left(\alpha,\beta\right)-N\left(\alpha,\beta^{\perp}\right)-N\left(\alpha^{\perp},\beta\right)+N\left(\alpha^{\perp},\beta^{\perp}\right)}{N\left(\alpha,\beta\right)+N\left(\alpha,\beta^{\perp}\right)+N\left(\alpha^{\perp},\beta\right)+N\left(\alpha^{\perp},\beta^{\perp}\right)}
	\end{aligned}
\end{equation}
which can be calculated from the coincidence rates, $N$, at different polarization angles. Here, we define additional angles, $\alpha^{\perp}=\alpha+90^o$ and $\beta^{\perp}=\beta+90^o$.
We chose $\alpha=0^o$, $\beta=22.5^o$, $\gamma=45^o$ and $\delta=67.5^o$ and confirmed that the inequality is strongly violated at $S=2.75\pm0.03$ by 25 standard deviations, which also agrees nicely with the expected value from the average visibility $S_{exp}=2\sqrt{2}V_{avg}\approx2.75$, indicating the quantum entanglement \cite{Peruzzo_2023,PhysRevLett.23.880}.
\begin{figure*}[htbp!]
\centering
\includegraphics[width=\textwidth]{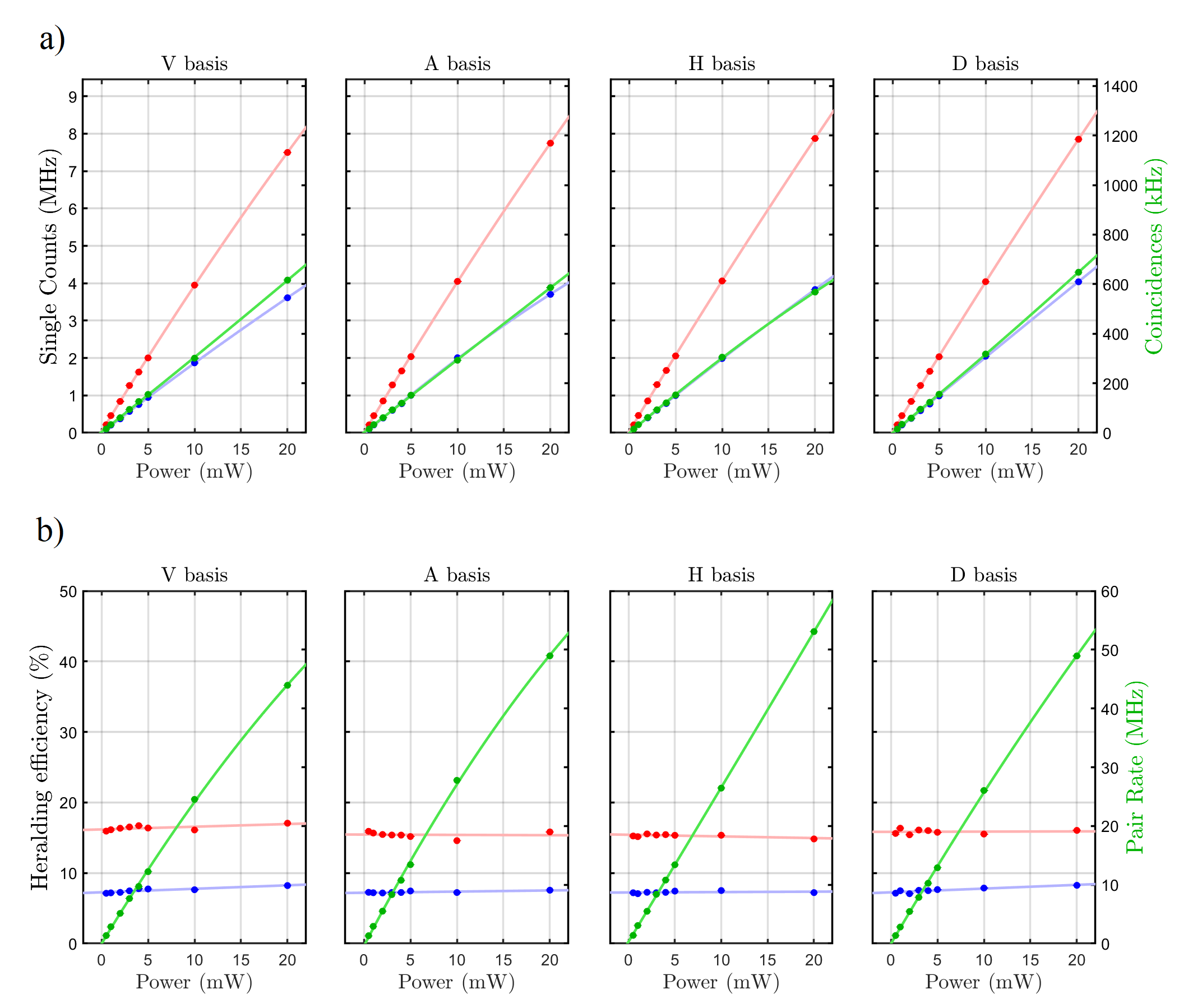}
\footnotesize\caption{(a) The single counts of both signal (red) and idler (blue) photons, as well as the coincidence (green) counts measured from four different polarization basis while varying the pump power. (b) The photon pair generation rates $R_{pair}$ (green), along with the heralding efficiencies of the signal photon $\eta_s$ (blue) and the idler photon $\eta_i$ (red) for different pump powers, calculated from (a).}\label{PowerVariation1}
\end{figure*}

We also examine the consistency of experimental data quality with increasing pump power levels. The experimental data includes the single counts of signal($N_s$) and idler($N_i$) photons, as well as the coincidence counts($N$). The results are plotted in Figure \ref{PowerVariation1}a. At a pump power of $\SI{1.0}{\milli\watt}$, the average background-subtracted values for four different bases were $N_s=(460.7\pm3.5)$kHz, $N_i=(210.7\pm1.0)$kHz and $N=(33.33\pm0.05)$kHz, which correspond to the maximum counts shown in Figure \ref{UltimateResult}. The calculated pair rates and the heralding efficiencies are plotted in Figure \ref{PowerVariation1}b. The pair rate from the V-basis at $\SI{1.0}{\milli\watt}$ was $R_{pair}=(2.92\pm0.12)$MHz.  Also, signal and idler heralding efficiencies were $\eta_s=15.8\%$ and $\eta_i=7.2\%$, respectively. The pair rate from all four bases show a linear increase with increasing pump power, while the heralding efficiencies remain fairly constant, as one would expect.
\section{Discussion and Outlook}

As shown in this experiment, entanglement of highly non-degenerate photon pairs could suffer from not only temporal walk-offs but also non-negligible spatial walk-offs. Our novel compensation method offers an effective and simple solution for compensating both the spatial and temporal walk-off of photon fields at any wavelength. With this compensation method, the coincidence rate at $\SI{1.0}{\milli\watt}$ of pump power was $N=(33.33\pm0.05)$kHz, showing a significant improvement from using the same setup without compensation, which was $N=(4.20\pm0.01)$kHz. At the same power, the average entanglement visibility of $(97.1\pm0.3)\%$, and the average pair generation rates of $(2.92\pm0.12)$MHz from all four bases were observed.\\
Extrapolating from the data shown in Figure \ref{PowerVariation1}b, $\SI{100}{\mega\hertz}$ of pair rate could possibly be achieved with a high-resolution time-tagging unit at the power level of only around $\SI{50}{\milli\watt}$, making it a promising source for long distance QKD \cite{Bourgoin_2013}. The setup could be simplified by adjusting the length of the second beam displacer so that either the signal or idler beam experiences no spatial walk-off, and then using birefringent wedges to compensate for the remaining spatial walk-off of the other beam, as described in this paper. With the validation from our experiment, we anticipate this compensation method contributing not only to advancements in long-distance quantum networks but also to other practical quantum entanglement applications that require non-degenerate entangled photon pairs.

\section{Acknowledgement}

The authors acknowledge funding from the Natural Sciences and Engineering Research Council of Canada (NSERC), Canada Foundation for Innovation (CFI), Ontario Research Fund (ORF) and Canadian Space Agency (CSA). We thank Younseok Lee for assistance with the purchases of PPLN crystal and the CW laser. We thank Ramy Tannous and Henri Morin for assistance with handling equipments in the lab.

\newpage
\bibliography{SungeunOh1D4.bib}

\end{document}